\documentclass[conference]{IEEEtran}
\IEEEoverridecommandlockouts
\usepackage[square,numbers]{natbib}
\usepackage{amsmath,amssymb,amsfonts}
\usepackage{algorithmic}
\usepackage{graphicx}
\usepackage{textcomp}
\usepackage{xcolor}
\usepackage{multirow}
\DeclareGraphicsExtensions{.png}

\begin{document}
\title{Towards a Security Cost Model for\\Cyber-Physical Systems}

\author{
	\IEEEauthorblockN{Igor Ivkic$^{1, 2}$, Andreas Mauthe$^{1}$, and Markus Tauber$^{2}$}
	\\
	\IEEEauthorblockA{$^1$\textit{Lancaster University}  - Lancaster, UK
		\IEEEauthorblockA{$^2$\textit{University of Applied Sciences Burgenland} - Eisenstadt, Austria}}	
}

\maketitle

\begin{abstract}
	In times of Industry 4.0 and cyber-physical systems (CPS) providing security is one of the biggest challenges. A cyber attack launched at a CPS poses a huge threat, since a security incident may affect both the cyber and the physical world. Since CPS are very flexible systems, which are capable of adapting to environmental changes, it is important to keep an overview of the resulting costs of providing security. However, research regarding CPS currently focuses more on engineering secure systems and does not satisfactorily provide approaches for evaluating the resulting costs. This paper presents an interaction-based model for evaluating security costs in a CPS. Furthermore, the paper demonstrates in a use case driven study, how this approach could be used to model the resulting costs for guaranteeing security. \\
\end{abstract}


\section{Introduction}
Industry 4.0 is driven by cyber-physical systems (CPS) and the internet of things (IoT) \cite{ref01, ref02}, where computation, communication and control functions integrate the cyber and physical worlds \cite{ref03, ref04}. A CPS consists of interconnected IoT-devices, sensors and actuators, which are capable of measuring the physical environment, analysing it and guiding intelligent actions to affect it. The IoT can be considered as the backbone of a CPS, which connects this swarm of IoT-devices, sensors and actuators \cite{ref05}. Unfortunately, advanced persistent threats (APT) \cite{ref06} launched at a CPS can cause disruptions transcending both the cyber realm and affecting the physical world \cite{ref07}. Stuxnet is just one example of such an attack, where several complex techniques have been used to interrupt the Iranian nuclear program \cite{ref08}. This case perfectly demonstrates how a simple malware attack on a CPS can have catastrophic consequences in the physical world, leading to the emergence of numerous new security challenges \cite{ref09}.

One of these challenges is that the sheer number of IoT-devices, sensors and actuators within a CPS need to be managed. Similar to mobile device management (MDM), it is necessary to control the way how new IoT-devices enter a CPS, and how they interact with each other within it. One way of meeting this challenge could be using an IoT-framework for secure applications \cite{ref10}, where devices, systems and services can be managed in a local cloud environment. Additionally, the framework should provide a procedure, which ensures that only valid and authorized IoT-devices can host software (SW) systems and services within the local cloud \cite{ref11}.

\begin{figure}[h] \centering \includegraphics[width=\columnwidth]{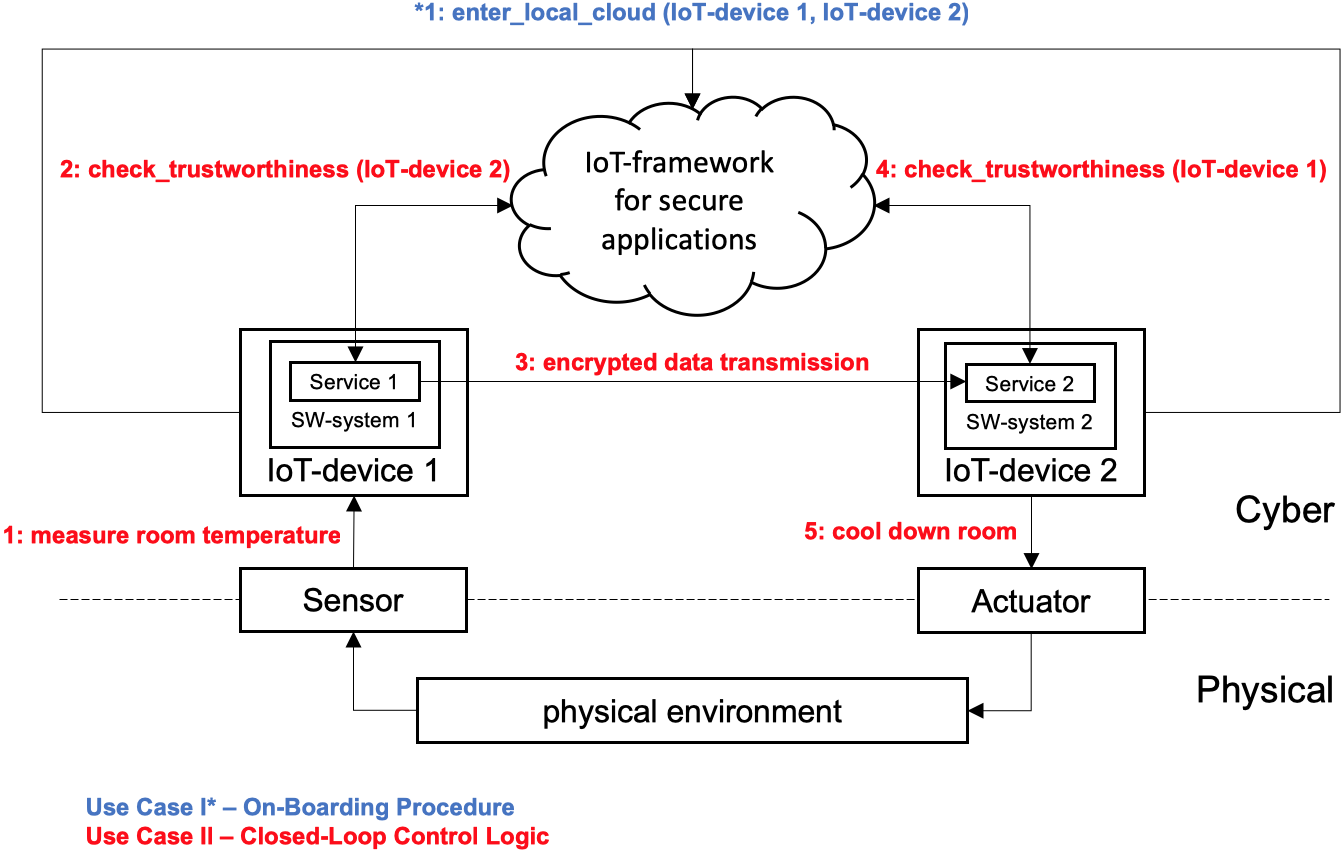} \caption{Overall use cases: entering a CPS and regulating a room's temperature} \end{figure} 

Another challenge is that IoT-devices, after they have been successfully enrolled to be part of a CPS, have to perform additional security tasks when interacting with each other. For instance, as shown in Fig. 1, IoT-device 1 uses a sensor to measure the physical environment (e.g. a room's temperature). Before the measured data is sent to IoT-device 2, additional security-related tasks need to be performed. First, it has to be verified whether the receiving device (IoT-device 2) is part of the same local cloud, and second, the temperature data needs to be encrypted before the transmission. This closed-loop CPS \cite{ref12} guarantees that both IoT-devices are trustworthy (because they are part of the same local cloud) and the communication is secure (because of the encryption).

Even though performing additional steps for guaranteeing security are essential, they always require compromises. For instance, it takes a certain amount of time to verify whether IoT-device 2 is trustworthy, and to encrypt the temperature data before transmission. Hence, in other words, the additional time it takes to guarantee security (e.g. encryption) can be considered as the cost of security. Due to the complexity of a CPS and its interconnected IoT-devices, sensors and actuators, an approach is needed for modelling security costs in CPS.

In this paper, we present an interaction-based model for evaluating security costs for CPS. First, we define the elements of a CPS and explain the basic principles of our approach. Next, we describe two different use cases and show in an evaluation how the presented approach can be used to model the security costs of these use cases. Our contribution towards a security cost model for CPS is twofold:
\begin{itemize}
	\item firstly, we introduce a mathematical expression for aggregating the effort it takes to provide security for CPS and 
	\item secondly, we demonstrate how this expression can be used to describe the security costs in two different use cases. \\
\end{itemize}

The remainder of this paper is organized as follows: Section II summarizes the related work in the field and points out the background of this paper. Next, in Section III, we define the elements of a CPS and introduce an interaction-based mathematical expression for evaluating security costs. Finally, in Section IV, we describe two use cases and show in an exemplary evaluation how to model the security costs of the presented use cases.

\section{Related Work and Background}
Measuring cyber security has been subject to many studies, resulting in proposing frameworks, methods and metrics for evaluating the security of specific systems, without referring to the resulting costs of security. In addition to that some of the presented approaches are limited by the usage of a single metric. For instance, in \cite{ref13, ref14} the authors show how the performance of a process can be measured, while the focus in \cite{ref15, ref16, ref17, ref18} is on evaluating the energy consumption. Other related work use methods and frameworks to evaluate how secure e.g. a system is \cite{ref19, ref20, ref21, ref22, ref23}. In other words, these solutions evaluate whether a security control has been implemented or not, instead of measuring the actual costs. A summary of related work regarding security metrics has been provided by Yee in \cite{ref24}. He first establishes the argument that many security metrics exist, but most of them are ineffective and not meaningful. Next, he provides a definition of a "good" and a "bad" metric and explains the difference between "traditional" and "scientifically based" security metrics. Finally, Yee presents his literature search on security metrics, which is based on various frameworks. 

The vast majority of research related to Industry 4.0 and CPS currently focuses on general challenges \cite{ref25, ref05}, design principles \cite{ref01, ref04}, or engineering \cite{ref02, ref03}. However, in \cite{ref09} the authors give an overview of the security concerns in CPS, identify challenges and summarize countermeasures. Rajkumar, De Niz and Klein \cite{ref12} demonstrate further that the complexity of CPS requires more effort to analyse and defend it. The reason for that is the explosion of states when considering combinations of events. Additionally they provide theoretical approaches for dealing with cyber threats and countermeasures models for CPS under attack. Other related work provides a local cloud environment (Arrowhead Framework) for managing the swarm of IoT-devices, sensors and actuators based on a service-oriented architecture (SOA) \cite{ref10}. Nevertheless, in comparison to the identified related work, this paper provides a more general, interaction-based approach for evaluating security costs in CPS. 

This paper is a continuation of \cite{ref26} where we introduced a high-level process flow based on Six Sigma for identifying, categorizing, analysing and eliminating security risks and measuring the resulting costs. This initial investigation included the evaluation of (i) how security risks of a smart business use case can be eliminated by implementing security controls, and (ii) how the resulting costs could be measured using a monetary cost metric (Euro). Even though the two use cases used IoT-devices and cloud computing the evaluation did not include the costs of security for a CPS. To extend this work the key new contribution of this paper is to present a mathematical expression, which can be used to describe and evaluate security costs of a CPS and which allows the usage of more then one cost metric.

\section{Modelling Security Costs}
In this section we present an approach for evaluating the costs of security in CPS. We first present an onion layer model and explain the general functionality of the approach. Then, we describe how it can be used to evaluate security-related tasks performed by components, which participate in an interaction of a CPS. Finally, we discuss possible outcomes including the cause and effect of evaluating the cost of security by using the presented model. 

\subsection{Definition}
In many respects, the control system view in Fig. 1 corresponds to the most fundamental definition of what CPS are. However, this closed-loop control system view might unintentionally give the impression that a CPS consists of a single control logic, which uses many sensors and actuators to interact with the physical environment. In reality, a CPS is defined by a set of components, which interact with each other and the physical environment. These components refer to any hardware or software resources that are capable of computation, communication and controlling sensors and/or actuators. All in all, this extended definition allows describing even more complex CPS (such as the CPS in Fig. 1). 

Within a CPS components interact with each other to serve a specific purpose. An interaction is a unit of work performed within a CPS, and treated in a coherent and reliable way independent of other interactions. Furthermore, an interaction refers to any usage of sensors, control functions and actuators, is executed at a specific point in time, and includes one or more participating components. Each component performs a number of different tasks, which can either be ordinary or security-related. An ordinary task, on the one hand, could be using a sensor to measure the physical environment, or using an actuator to change it. A security-related task, on the other hand, could be encrypting the measured data before sending it to another component. Thus, it is important to differentiate between those two types, since an evaluation of security costs should only focus on security-related tasks. 

\subsection{Onion Layer Model}
As previously defined, the security costs of a CPS including its interactions, components and security-related tasks could be modelled by using an onion layered approach. As shown in Fig. 2, the entirety of security costs in a CPS results in using one ore more metrics to measure the total amount of performed security-related tasks. Furthermore, the sum of all security-related tasks is performed by the total number of components within a CPS, which in turn can be part of various interactions. Summarizing, to evaluate how much it costs to guarantee security in a CPS, including all layers of Fig. 2, the onion layer model propose to form a sum of sums. 

\begin{figure}[h] \centering \includegraphics[width=\columnwidth]{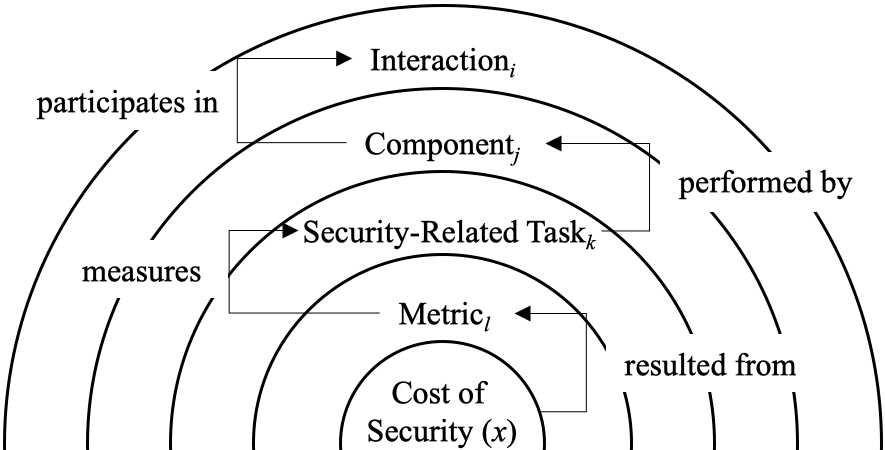} \caption{Onion layer model for evaluating security costs in CPS} \end{figure} 

The presented onion layer approach describes security costs that are incurred each time an interaction is performed in a CPS. As defined in (1), the mathematical expression describes the total security costs of a CPS as a result of four different summations. The first sum ($\sum_{i=1}^{n}$) represents all existing interactions of a CPS, while the second ($\sum_{j=1}^{m}$) summarizes all components within one interaction. Followed by that is the sum ($\sum_{k=1}^{o}$) of all security-related tasks, which have been performed by a specific component. Finally, the last sum ($\sum_{l=1}^{p}$) adds up all metrics used to measure the performance of a specific security-related task. This mathematical expression enables to create a sum of sums, which includes all interactions, components, security-related tasks and metrics used to evaluate the security costs of a CPS.

\begin{equation}
	\sum\limits_{i=1}^{n} \sum\limits_{j=1}^{m} \sum\limits_{k=1}^{o} \sum\limits_{l=1}^{p} x_{ijkl}, \indent \indent (i, j, k, l, n, m, o, p \in \mathbb{N})
\end{equation}\\

The mathematical expression in (1) can be used in different ways to produce different kinds of results. For instance, a set of metrics could be used to evaluate the costs of performing security-related tasks, which are being performed by interconnected components of an interaction. In addition to that the measured security-related costs could be represented by either producing an overall result (total costs of an interaction), or by aggregating the results on different levels (e.g. per interaction, per component, per security-related task, per metric). Table I shows an example of how the costs of security could be aggregated (on different layers) by using two different metrics for measuring a various set of security-related tasks, which have been performed by two different components of one interaction:

\begin{table}[htbp]
	\caption{Example I: Aggregation of Security Costs in a CPS}
	\begin{center}
		\begin{tabular}{|c|c|c|c|c|}
			\hline
			\multicolumn{5}{|c|}{\textbf{Interaction}} \\ \hline
			\multirow{2}{*}{\textbf{Components}} & \multirow{2}{*}{\textbf{Tasks}} & \multicolumn{2}{c|}{\textbf{Metrics}} & \multirow{2}{*}{\textbf{$\sum$}} \\ \cline{3-4}
			& & {\textbf{1}} & {\textbf{2}} &  \\ \hline
			\multirow{3}{*}{\textbf{1}}		& \textbf{1} & \textbf{$x_{111}$} & \textbf{$x_{112}$} & \textbf{$\sum\limits_{l=1}^{1} x_{1l}$} \\ \cline{2-5}
			& $\sum$ & \textbf{$\sum\limits_{k=1}^{1} x_{k1}$} & \textbf{$\sum\limits_{k=1}^{1} x_{k2}$} & \textbf{$\sum\limits_{k=1}^{1} \sum\limits_{l=1}^{2} x_{kl}$} \\ \hline
			\multirow{4}{*}{\textbf{2}}		& \textbf{1} & \textbf{$x_{211}$} & \textbf{$x_{212}$} & \textbf{$\sum\limits_{l=1}^{2} x_{1l}$} \\ \cline{2-5}
			& \textbf{2} & \textbf{$x_{221}$} & \textbf{$x_{222}$} & \textbf{$\sum\limits_{l=1}^{2} x_{2l}$}  \\ \cline{2-5}
			& $\sum$ & \textbf{$\sum\limits_{k=1}^{2} x_{k1}$} & \textbf{$\sum\limits_{k=1}^{2} x_{k2}$} & \textbf{$\sum\limits_{k=1}^{2} \sum\limits_{l=1}^{2} x_{kl}$} \\ \hline
		\end{tabular}
		\label{tab1}
	\end{center}
\end{table}

\subsection{Normalization}
Evaluating security costs of a CPS a promising approach for predicting coming events and/or influencing future decisions. In order to be able to adapt to environmental changes, a CPS applies policies on component level. For instance, a self-adaptability policy could be applied to start an additional component, when more computational resources are needed to perform certain tasks (e.g. more computational power). However, in terms of security these policies can be used to either provide security to a CPS or restore it after an attack. Now, past measurement data of security costs of a CPS could therefore be used to implement a self-adaptation policy, which aims at providing the same level of security within a CPS, but at lower costs.

Even though, the presented approach can be used to evaluate the security costs of all interactions of a CPS, it implies using metrics with aggregatable measurment results. In other words, a metric might provide results, which cannot be aggregated with the results provided by another metric, due to incompatible units. Thus, the duration given in milliseconds (ms) uses a different unit than the load of a central processing unit (CPU) given in percent. Another problem is that when using two or more metrics with different units the results my require interpretation in order to make sense. For example, measuring the duration of a security task and the CPU load used to do so might provide the following two results: 
\begin{itemize}
	\item \textit{x$_{1}$} = 5 ms + 10\%
	\item \textit{x$_{2}$} = 10 ms + 5\%
	\\
\end{itemize}

Without normalization measured data provided by different metrics it is impossible to tell, which of the two measurements is \text{"better" or "cheaper"} in terms of security costs (e.g. \textit{x$_1$ $<$ \textit{x$_2$}} or \textit{x$_1$} $>$ \textit{x$_2$}). The presented mathematical expression in (1) assumes in general that the results provided by the used metrics can be aggregated. Normalizing results provided by metrics with different units will not be further elaborated here (future work). 

Another point is that the onion layer model only considers dependencies between layers in terms of security costs. For instance, two consecutive tasks (task 1 and task 2) may not be performed every time. Instead task 2 could be only executed, when a specific condition is met in task 1. Now, depending on how many tasks have been performed the dependency relationship between task 1 and task 2 will only be shown in the different cost measurements. A dependency analysis will be subject of future work and will not be further elaborated in this paper. 

\section{Use Case Description and Evaluation}
In the previous section we have defined that a CPS consists of different components, which combine computation, communication and controlling abilities to interact with other components and the physical world. Additionally, we described an onion layer model, which can be used to evaluate security costs of a CPS. In this section we describe two different use cases and demonstrate how the approach from Section III can be used to evaluate them. These two use cases are based on the CPS from Fig. 1, which consists of three components that interact with each other and with the physical environment. The first component (IoT-device 1) uses a sensor to measure the temperature of a room. Next, it uses an IoT-framework to verify whether the second component (IoT-device 2) is trustworthy, before transmitting the data. The same verification is done on the receiving side, where the IoT-device 2 verifies whether IoT-device 1 is trustworthy. Finally, after the IoT-framework confirms the trustworthiness of both IoT-devices, the second device starts cooling down the room, but only if the temperature is over a certain limit (e.g. 25 degree Celsius). Summarizing, this CPS uses three different component to measure the temperature of a room, perform security-related tasks to guarantee trustworthiness between components and effectively cool down the room (if the limit has been reached). 

\subsection{Use Case I: On-Boarding Procedure}
As previously mentioned, a CPS is formed by its components, which are capable of interacting with each other and with the physical world. Although, before these components start interacting with each other, they first need to be "on-boarded" (or enrolled) to become part of a CPS. In other words, the IoT-devices need to register with an IoT-framework including their SW-system and produced services. For instance, IoT-device 1 uses SW-system 1 to control a temperature sensor and produces the service "measure room temperature". One way of enabling this on-boarding procedure is by using the Arrowhead framework \cite{ref10}, which can be used to create a SOA local cloud environment for managing the swarms of components, sensors and actuators within a CPS. 

This Arrowhead local cloud provides an on-boarding procedure, which controls the way how an IoT-device including its SW-system and produced service enters a CPS. Furthermore, as Bicaku et al. \cite{ref11} explained, it establishes a chain of trust to assure that the Arrowhead local cloud is not compromised upon the introduction of a new component. This is crucial, from a security perspective, since the on-boarding procedure guarantees that each component passed all requirements before being allowed to enter the Arrowhead local cloud. Additionally, it creates a trustworthy environment, where a component can use the Arrowhead cloud systems to verify whether another component is part of the same cloud before starting an interaction.

So, before the two IoT-devices from Fig. 1 start controlling the temperature of a room, they first have to enter the Arrowhead local cloud. By doing so, they have to go through the required steps of the on-boarding procedure in order to successfully enter the CPS. As explained in \cite{ref11}, the Arrowhead local cloud is composed of a number of systems, which perform specific tasks in the on-boarding procedure. Only if a new device successfully passes all tasks of all these systems it is allowed to enter the local cloud. Now, to support the overall use case from Fig. 1 there are two IoT-devices, which need to complete the on-boarding procedure. In other words, for each IoT-device there is an interaction between the device and the Arrowhead local cloud. During an interaction different tasks are performed by either the IoT-device or by one of the Arrowhead systems. Some of these tasks can be relevant to security, where e.g. a certificate is being transmitted and checked for validity. Finally, after both interactions have been completed successfully, the two IoT-devices are part of the CPS and can start regulating the room's temperature.

\subsection{Evaluation of Use Case I}
The on-boarding procedure is used by both IoT-devices (from Fig. 1) to enter the CPS. In other words, to "on-board" both devices the same procedure needs to be completed twice, but the steps for each run are identical. Therefore only one interaction needs to be considered when evaluating the costs of security for use case I. However, it is necessary to separate ordinary and security-related tasks and to assign them to the performing component. According to the authors of \cite{ref11} the following have been identified as security-related tasks:
\begin{itemize}
	\item \textbf{task 3:} Publish(DeviceRecord, device-key) \\ performed by: IoT-device 1 and IoT-device 2 
	\item \textbf{task 4:} Authenticate(AuthenticationRequest) \\ performed by: Arrowhead local cloud
	\item \textbf{task 5:} Publish(SystemRecord, SW-key, local-cloud-SW-key) \\ performed by: IoT-device 1 and IoT-device 2 	
	\item \textbf{task 6:} Authorise(AuthorisationRequest) \\ performed by: Arrowhead local cloud
	\item \textbf{task 7:} Publish(ServiceRecord) \\ performed by: IoT-device 1 and IoT-device 2
	\item \textbf{task 8:} Authorise(AuthorisationRequest) \\ performed by: Arrowhead local cloud\\
\end{itemize} 

The execution of the on-boarding procedure for the first IoT-device means that the two components (IoT-device 1 and IoT-framework) perform a total of 6 security-related tasks (the same applies for on-boarding IoT-device 2). As shown in (2) and (3) the onion layer model can be used to describe the \begin{figure*} \centering \includegraphics[width=\textwidth]{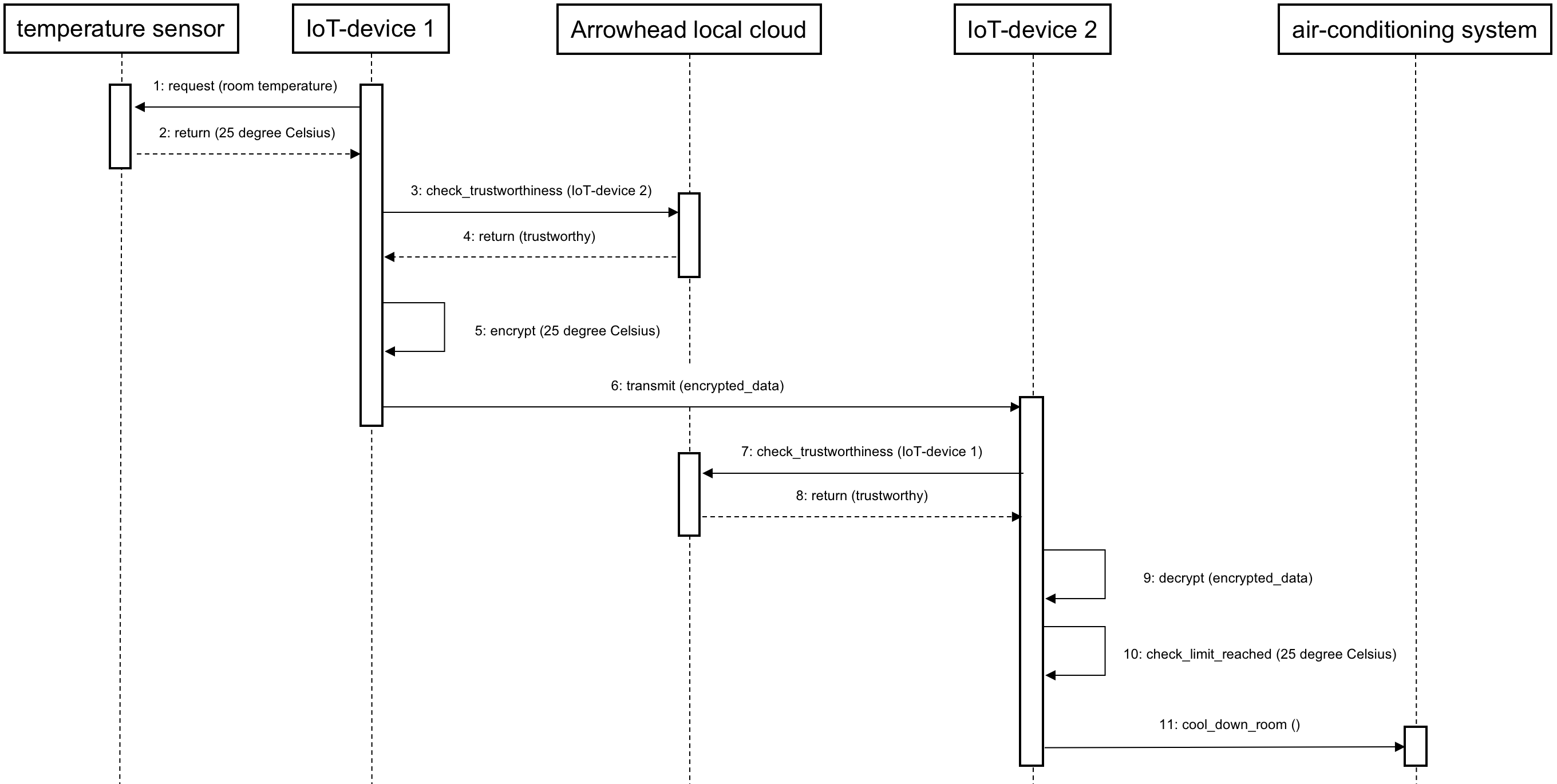} \caption{The sequence diagram for the overall use case for controlling the room temperature} \end{figure*}

\noindent resulting costs (\textit{c$_1$, c$_2$}) for on-boarding both IoT-devices by using an exemplary metric for measuring the duration in ms:

\begin{equation}
	c_1 = \sum\limits_{i=1}^{1} \sum\limits_{j=1}^{2} \sum\limits_{k=1}^{3} \sum\limits_{l=1}^{1} x_{ijkl}
\end{equation}\\
\begin{equation}
	c_2 = \sum\limits_{i=1}^{1} \sum\limits_{j=1}^{2} \sum\limits_{k=1}^{3} \sum\limits_{l=1}^{1} x_{ijkl}
\end{equation}\\

The following table shows a summarized view on the evaluation of security costs of the on-boarding procedure of both IoT-devices:
\begin{table}[htbp]
	\caption{On-Boarding Precedure Security Cost Evaluation}
	\begin{center}
		\begin{tabular}{|c|c|c|c|}
			\hline
			\multicolumn{4}{|c|}{\textbf{Interaction 1}} \\ \hline
			\textbf{Component} & \textbf{Task} & \textbf{Duration} & \textbf{$\sum$} \\ \hline 
			\multirow{3}{*}{\textbf{1$^{\mathrm{a}}$}}		& \textbf{3} & \text{3 ms} & \multirow{3}{*}{\text{17 ms}} \\ \cline{2-3}
			& {\textbf{5}} & \text{5 ms} & \\ \cline{2-3}
			& {\textbf{7}} & \text{9 ms} & \\ \cline{2-3} \hline
			\multirow{3}{*}{\textbf{2$^{\mathrm{b}}$}}		& \textbf{3} & \text{3 ms} & \multirow{3}{*}{\text{17 ms}} \\ \cline{2-3}
			& {\textbf{5}} & \text{5 ms} & \\ \cline{2-3}
			& {\textbf{7}} & \text{9 ms} & \\ \cline{2-3} \hline
			\multirow{3}{*}{\textbf{3$^{\mathrm{c}}$}}		& \textbf{4} & \text{8 ms} & \multirow{3}{*}{\text{26 ms}} \\ \cline{2-3}
			& {\textbf{6}} & \text{14 ms} & \\ \cline{2-3}
			& {\textbf{8}} & \text{4 ms} & \\ \cline{2-3} \hline
			\multicolumn{4}{r}{\text{$^{\mathrm{a}}$IoT-device 1; $^{\mathrm{b}}$IoT-device 2; $^{\mathrm{c}}$Arrowhead local cloud}}
		\end{tabular}
		\label{tab2}
	\end{center}
\end{table}	\vspace{-5mm}

\subsection{Use Case II: Closed-Loop Temperature Control}
Once the on-boarding procedure has been completed successfully and the two IoT-devices are part of the CPS they can start regulating the room's temperature. In order to do so, each component performs different tasks and interacts with other components. First, IoT-device 1 uses a sensor to measure the room’s current temperature. Then, the first security-related task is performed by asking the Arrowhead local cloud, whether IoT-device 2 is trustworthy or not. The trustworthiness is verified by the systems of the local cloud, by checking if IoT-device 2 has successfully passed the on-boarding procedure. After the check confirms that IoT-device 2 is part of the Arrowhead local cloud, the first IoT-device performs another security-related task. This time an encryption algorithm is used to cipher the previously measured data, before it is being transmitted to the second IoT-device.

However, before IoT-device 2 decrypts the received message, it first uses the Arrowhead local cloud to verify, whether IoT-device 1 is trustworthy or not. If it is, the message is decrypted and checked, if the measured temperature is over a predefined limit (e.g.: 25 degree Celsius). Only if this limit has been reached, the second IoT-device sends a command to an air-conditioning system to cool down the room. Fig. 2 shows this closed-loop control logic from measuring the room’s temperature by using a sensor to performing security-relevant tasks and finally using an actuator to change the temperature.

\subsection{Evaluation of Use Case II}
The complexity is even greater in the second use case, due to the interconnection of all three components to reach a common goal. This includes measuring the temperature of a room, verifying the trustworthiness of the two IoT-devices, establishing an encrypted data transmission and finally cooling down the room. To reach this goal, each component performs different tasks, which again need to be divided into ordinary and security-related tasks and then assigned to the performing component. As shown in Fig. 2, the following have been identified as security-related tasks:
\begin{itemize}
	\item \textbf{task 3:} check\_trustworthiness(IoT-device 2) \\ performed by: IoT-device 1
	\item \textbf{task 4:} return(trustworthy) \\ performed by: Arrowhead local cloud
	\item \textbf{task 5:} encrypt(25 degree Celsius) \\ performed by: IoT-device 1
	\item \textbf{task 7:} check\_trustworthiness(IoT-device 1) \\ performed by: IoT-device 2
	\item \textbf{task 8:} return(trustworthy) \\ performed by: Arrowhead local cloud
	\item \textbf{task 9:} decrypt(encrypted\_data) \\ performed by: IoT-device 2\\
\end{itemize} 

The execution of the closed-loop temperature control involves three components (IoT-device 1, IoT-device 2, Arrowhead local cloud), which perform a total of 6 security-related tasks. Again, as shown in (4), the resulting costs (\textit{c$_3$}) can be described by the onion layer model and evaluated by using the same metric as used in the previous use case evaluation (duration measured in ms):
\begin{equation}
	c_3 = \sum\limits_{i=1}^{1} \sum\limits_{j=1}^{3} \sum\limits_{k=1}^{2} \sum\limits_{l=1}^{1} x_{ijkl}
\end{equation}\\ \vspace{-5mm}

The following table summarizes the evaluation of security costs of the closed-loop temperature control:
\begin{table}[htbp]
	\caption{Closed-Loop Temperature Control Security Cost Evaluation}
	\begin{center}
		\begin{tabular}{|c|c|c|c|}
			\hline
			\multicolumn{4}{|c|}{\textbf{Interaction 2}} \\ \hline
			\textbf{Component} & \textbf{Task} & \textbf{Duration} & \textbf{$\sum$} \\ \hline 
			\multirow{2}{*}{\textbf{1$^{\mathrm{a}}$}}		& \textbf{3} & \text{2 ms} & \multirow{3}{*}{\text{10 ms}} \\ \cline{2-3}
			& {\textbf{5}} & \text{8 ms} & \\ \cline{2-3} \hline
			\multirow{2}{*}{\textbf{2$^{\mathrm{b}}$}}		& \textbf{7} & \text{2 ms} & \multirow{3}{*}{\text{9 ms}} \\ \cline{2-3}
			& {\textbf{9}} & \text{7 ms} & \\ \cline{2-3} \hline
			\multirow{2}{*}{\textbf{3$^{\mathrm{c}}$}}		& \textbf{4} & \text{1 ms} & \multirow{3}{*}{\text{2 ms}} \\ \cline{2-3}
			& {\textbf{8}} & \text{1 ms} & \\ \cline{2-3} \hline
			\multicolumn{4}{r}{\text{$^{\mathrm{a}}$IoT-device 1; $^{\mathrm{b}}$IoT-device 2; $^{\mathrm{c}}$Arrowhead local cloud }} \\
		\end{tabular}
		\label{tab2}
	\end{center}
\end{table}

The results in Table III perfectly demonstrate why it is important to evaluate security costs in CPS. For instance, if the same evaluation was done periodically (e.g. measuring the room's temperature every 5 minutes) the results would show that the security costs remain constant regardless of the temperature limit being reached or not. As shown in Fig. 3, only after all security-related task (task 3 to task 8) have been performed it is verified, whether the temperature limit has been reached or not. So, the security costs could be significantly reduced in this CPS, if IoT-device 1 performs step 10 (check\_limit\_reached) after receiving the measured data from the sensor (step 2) and before checking the trustworthiness of IoT-device 2 (step 3). By doing so, the security-related tasks would only be performed when the temperature limit has been reached, which would reduce the security costs immensely.

\section{Conclusion and Future Work}
In this paper, we introduced a new approach for evaluating security costs in CPS. We presented an overall use case including the interaction of components, sensors and actuators in a CPS. In this regard, we defined in Section III that a CPS is formed by its components, which are capable of computation, communication and control functions. Furthermore, we proposed an onion layer model, which forms a sum of sums of all interactions, components, security-related tasks performed and metrics used to measure security costs. Next, we described tow use cases where multiple components participate in an interaction and perform both ordinary and security-related tasks. 

The first use case points out the necessary steps for entering and existing CPS without compromising it (on-boarding procedure). The second use case combines the usage of components, sensors and actuators to control the temperature of a room (closed-loop temperature control). Finally, we demonstrate how the proposed onion layer model can be used to evaluate the security costs of the two use cases. In this demonstration we used an exemplary metric (duration measured in ms) to show how to evaluate the duration of performing all security-related tasks (measured in ms).  

The main contribution of this paper is the initial investigation of an approach for modelling security costs in CPS. This will be enhanced in future work by considering more use cases, providing a metric catalogue and exploring methodologies for normalizing measured data by different metrics. Furthermore, the mathematical expression presented in Section III (1) will be part of further research to create a more general, algorithmic and parametric equation for modelling security costs in CPS. Summarizing, the main goal is to develop a framework that uses metrics from catalogue to evaluate security costs, normalize the results and is applicable for any use case.

\section*{Acknowledgment}
Research leading to these results has received funding from the EU ECSEL Joint Undertaking under grant agreement n°737459 (project Productive4.0) and from the partners’ national programmes/funding authorities and the project MIT 4.0 (FE02), funded by IWB-EFRE 2014 - 2020 coordinated by Forschung Burgenland GmbH.

\bibliographystyle{unsrtnat}
\bibliography{references}

\end{document}